\documentclass[twocolumn,secnumarabic,amssymb, nobibnotes, aps, prl, superscriptaddress]{revtex4-2}

\usepackage{amsmath,mathrsfs,amssymb,amsfonts,amsthm,mathtools}
\usepackage{graphicx,color}
\usepackage{subcaption}
\usepackage{booktabs}
\usepackage{physics}
\usepackage[T1]{fontenc}
\usepackage{bm}
\usepackage{empheq}
\usepackage{hyperref}
\hypersetup{
    colorlinks=true,
    linkcolor=blue,
    citecolor=blue,
    urlcolor=blue
}

\newcommand{\nn}{\nonumber}
\newcommand*\widefbox[1]{\fbox{\hspace{2em}#1\hspace{2em}}}

\begin{document}
\title{Vacuum-Induced Quantum Gate}

\author{Arash Azizi}
\email{sazizi@tamu.edu}
\affiliation{The Institute for Quantum Science and Engineering, Texas A\&M University, College Station, TX 77843, U.S.A.}
\affiliation{Department of Physics and Astronomy, Texas A\&M University, College Station, TX 77843, U.S.A.}

\begin{abstract}
We demonstrate that the quantum vacuum, as perceived by a uniformly accelerating observer, can be harnessed to perform a quantum Z-gate. A two-level Unruh-DeWitt detector, prepared in a superposition of its ground and excited states, undergoes a second-order interaction with the vacuum, resulting in a two-photon emission. We derive the exact analytical form of the final entangled detector-field state and show that this emission is conditional on a phase flip of the detector's initial state—the defining feature of the gate's operation. This process harvests entanglement from the Minkowski vacuum, producing photon pairs entangled across causally disconnected Rindler wedges. This work reframes acceleration-induced radiation not as thermal noise but as a coherent computational resource, offering new pathways for relativistic quantum information.
\end{abstract}

\maketitle

\textit{Introduction.}---The Unruh effect posits that a uniformly accelerating observer perceives the Minkowski vacuum as a thermal bath at a temperature proportional to the acceleration \cite{Unruh1976, Fulling1973, Davies1975}. This observer-dependent phenomenon underscores the relativity of quantum field vacua \cite{Crispino2008, BirrellDavies1982, Takagi1986} and shares deep analogies with Hawking radiation from black holes \cite{Hawking1975, Wald:1975kc} and particle creation in expanding universes \cite{Parker1969quantized}. It has profound implications for quantum field theory in curved spacetimes \cite{HollandsWald2015} and relativistic quantum information (RQI) \cite{Peres2004RMP, Mann2012RQI, Alsing2012, Su2014communication, Foo2020teleportation, Tjoa2022teleportation}, where acceleration can degrade entanglement and other quantum resources \cite{FuentesSchuller2005BH, Bruschi2010QI_Unruh}.

The Unruh effect is commonly probed using the Unruh-DeWitt (UDW) detector model \cite{Unruh1976, Einstein100, Colosi2009Rovelli}---a localized two-level system that interacts with the quantum field and becomes excited by vacuum fluctuations, mimicking the detection of thermal radiation. While first-order perturbation theory reveals the thermal character of these excitations, higher-order processes uncover coherent, non-thermal structures in the vacuum \cite{Scully2022, Azizi2025TPE}, including entanglement across Rindler wedges \cite{Reznik2003, Svidzinsky21prr, Svidzinsky21prl}.

In this Letter, we demonstrate that the accelerating vacuum can serve as a computational resource, functioning as a quantum phase gate. A UDW detector prepared in a superposition of ground and excited states undergoes a second-order interaction, emitting two photons conditionally on a phase flip in its initial state. This process channels into distinct directional modes, producing entangled photon pairs spanning causally disconnected regions. In RQI, acceleration alters quantum tasks like teleportation \cite{Su2014communication, Foo2020teleportation, Tjoa2022teleportation} and computing \cite{Aspling2024UDW_QC, LeMaitre2025PRL}, often introducing noise. Here, we harness vacuum coherence to imprint information, opening pathways for relativistic protocols.

This vacuum-controlled Z-gate could be realized in analogs like circuit QED \cite{Nation2012Nori, Blais2021RevModPhys} or ion traps \cite{Alsing2005IonTrap, Lamata2011}, verified via Ramsey interferometry \cite{Polo-Gomez2022measurement}, and may extend to harvesting nonclassical resources like magic \cite{Nystrom2025PRR}.

\begin{figure*}[t]
    \centering
    \begin{subfigure}[b]{0.32\textwidth}
        \centering
        \includegraphics[width=\textwidth]{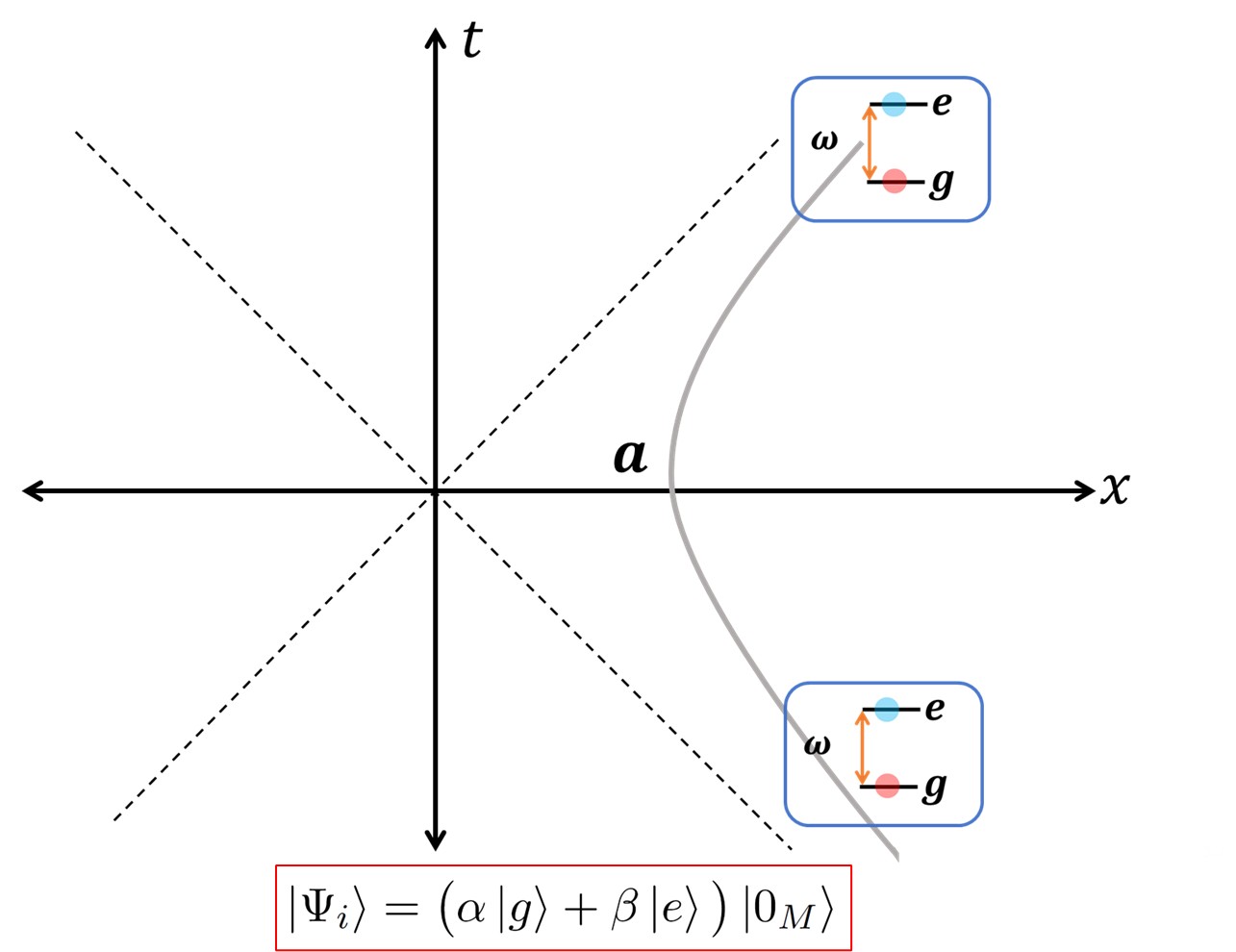} 
        \caption{The detector is prepared in a coherent superposition, erasing which-path information.}
        \label{fig:setup}
    \end{subfigure}
    \hfill
    \begin{subfigure}[b]{0.32\textwidth}
        \centering
        \includegraphics[width=\textwidth]{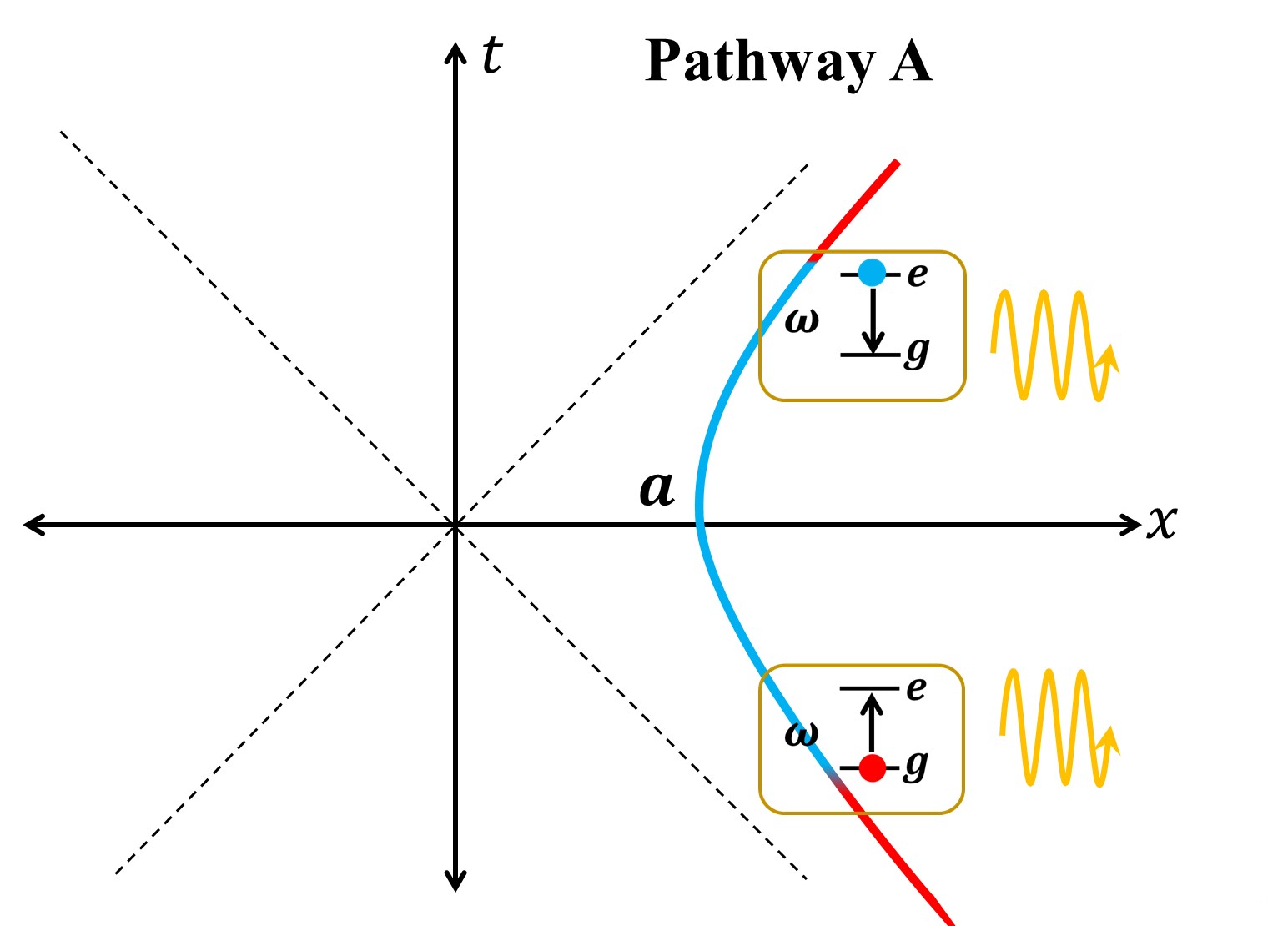} 
        \caption{Pathway A: The history for the $\ket{g}$ component (Unruh excitation then decay).}
        \label{fig:path_a}
    \end{subfigure}
    \hfill
    \begin{subfigure}[b]{0.32\textwidth}
        \centering
        \includegraphics[width=\textwidth]{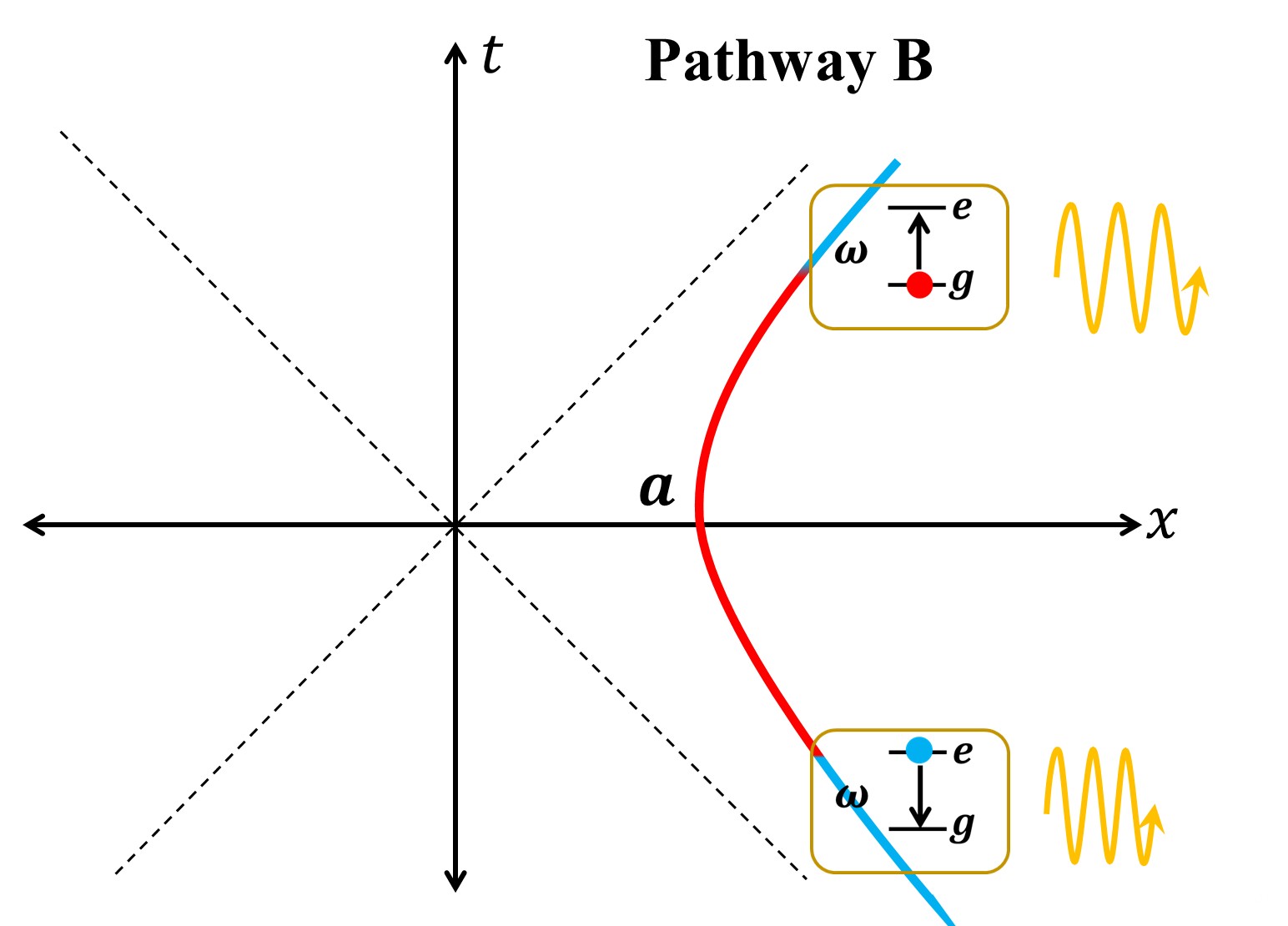} 
        \caption{Pathway B: The history for the $\ket{e}$ component (decay then Unruh excitation).}
        \label{fig:path_b}
    \end{subfigure}
    \caption{The setup for the relativistic quantum gate. When the detector is prepared in an initial superposition (a), the evolution proceeds along two indistinguishable quantum histories (b and c), whose interference determines the final state.}
    \label{fig:three_pathways}
\end{figure*}

\textit{Model.}---We consider a two-level UDW detector with energy gap $\omega$ coupled to a massless scalar field $\Phi$ in $(1+1)$-dimensional spacetime. The detector follows a hyperbolic worldline with uniform proper acceleration $a$. The interaction is governed by the Hamiltonian $H_{\text{int}} = g \frac{\partial }{\partial \tau} \Phi(\tau) (\sigma^{\dagger} e^{i \omega \tau} + \sigma e^{-i \omega \tau})$, where $\tau$ is the detector's proper time.

To analyze the interaction, we expand the field in the Unruh mode basis, which is natural for the accelerating frame (see Supplemental Material for details). This framework reveals a foundational non-local process. In the standard first-order Unruh effect, detector excitation ($\ket{g} \to \ket{e}$) is accompanied by the creation of a photon that is predominantly localized in the causally disconnected Rindler wedge. This non-local emission is a direct consequence of the entangled structure of the Minkowski vacuum and is foundational to the two-photon effect we analyze~\cite{UnruhWald1984}.

\textit{Interference of Spacetime Pathways.}---To realize the vacuum-induced gate, the detector is prepared in a coherent superposition of its internal states, as depicted in Fig.~1: $\ket{\Psi_{i}}=(\alpha\ket{g}+\beta\ket{e})\otimes\ket{0_{M}}$. The subsequent evolution is governed by the second-order Dyson series. Due to linearity, the final state is a superposition of two distinct quantum histories:
\begin{equation}
    \ket{\Psi_f} = \alpha \ket{\Psi_{g \to e \to g}} + \beta \ket{\Psi_{e \to g \to e}}.
\end{equation}
The first path, $\ket{\Psi_{g \to e \to g}}$, corresponds to the ground-state component undergoing Unruh excitation followed by radiative decay. The exact analytical form for this state has been derived previously~\cite{Azizi2025TPE} and is provided in the Supplemental Material for completeness.

The second path, $\ket{\Psi_{e \to g \to e}}$, corresponds to the excited-state component spontaneously decaying and then being re-excited. Crucially, the amplitude for this process can be obtained directly from the first through a fundamental symmetry. The only mathematical difference between the two pathways is the sign of the detector's frequency in the phase accumulated during its intermediate state ($e^{-i\omega(\tau-\tau')}$ vs. $e^{+i\omega(\tau-\tau')}$). This corresponds to the transformation $\omega \to -\omega$ on the final state amplitude. Applying this symmetry to the known result for $\ket{\Psi_{g \to e \to g}}$ yields the state for the counterpart process.

Superposing these two indistinguishable pathways yields the final entangled detector-field state. The explicit final state for the decay-first pathway, and the total superposed state are presented in Eqs.~\eqref{eq:ege} and \eqref{eq:total}.
\begin{widetext}
\begin{align}
 \ket{\Psi_{e \to g \to e}} &=
\frac{ -i g^2}{4\hbar^2}
\int_{-\infty}^{+\infty} d\Omega
\frac{1}{\sinh(\pi \Omega)}
\Bigg\{ \frac{\Omega}{\frac{\omega}{a} + \Omega}
 A^\dagger_{\Omega}  A^\dagger_{-\Omega}
 + \frac{\Omega}{\frac{\omega}{a} - \Omega}
 B^\dagger_{\Omega}  B^\dagger_{-\Omega}
 + \frac{2\frac{\omega}{a}\,\Omega\,  a^{2i\Omega} }{(\frac{\omega}{a})^2 - \Omega^2}
 A^\dagger_{\Omega}  B^\dagger_{\Omega} \Bigg\}
\ket{0_M} \ket{e}. \label{eq:ege} \\
\ket{\Psi_f} =& \frac{i g^2}{4\hbar^2} \int_{-\infty}^{+\infty} d\Omega \frac{1}{\sinh(\pi \Omega)} \Bigg\{
\alpha \Bigg( \frac{\Omega}{\frac{\omega}{a} - \Omega} A^\dagger_{\Omega} A^\dagger_{-\Omega} + \frac{\Omega}{\frac{\omega}{a} + \Omega} B^\dagger_{\Omega} B^\dagger_{-\Omega} + \frac{2\frac{\omega}{a}\,\Omega\, a^{2i\Omega} }{(\frac{\omega}{a})^2 - \Omega^2} A^\dagger_{\Omega} B^\dagger_{\Omega} \Bigg) \ket{g} \nn \\
&- \beta \Bigg( \frac{\Omega}{\frac{\omega}{a} + \Omega} A^\dagger_{\Omega} A^\dagger_{-\Omega} + \frac{\Omega}{\frac{\omega}{a} - \Omega} B^\dagger_{\Omega} B^\dagger_{-\Omega} + \frac{2\frac{\omega}{a}\,\Omega\, a^{2i\Omega} }{(\frac{\omega}{a})^2 - \Omega^2} A^\dagger_{\Omega} B^\dagger_{\Omega} \Bigg) \ket{e} \Bigg\}\otimes\ket{0_M}. \label{eq:total} 
\end{align}
\end{widetext}
The full expression for the final state, $\ket{\Psi_f}$, contains an integral over all Unruh frequencies $\Omega$. The physically dominant, on-shell contributions from this integral can be extracted at the resonance frequencies $\Omega = \pm \omega/a$ by applying the Sokhotski-Plemelj theorem. As detailed in the Supplemental Material, this procedure isolates the resonant part of the final state, $\ket{\Psi_f}_{\text{res}}$, which describes the real, energy-conserving two-photon emission.

This resonant state reveals the central result of our work: the two-photon emission process is conditional on a phase flip of the detector's initial qubit state. The final resonant state is given by:
\begin{widetext}
\begin{align}
\ket{\Psi_f}_{\text{res}} \approx \frac{\pi g^2}{4\hbar^2} \frac{\omega/a}{\sinh(\pi \omega/a)} \Bigg\{& A^\dagger_{\omega/a} A^\dagger_{-\omega/a}
+ B^\dagger_{\omega/a} B^\dagger_{-\omega/a} \nn\\
&
+ a^{2i\omega/a} A^\dagger_{\omega/a} B^\dagger_{\omega/a}
+ a^{-2i\omega/a} A^\dagger_{-\omega/a} B^\dagger_{-\omega/a} \Bigg\} \Big(\alpha\ket{g} - \beta\ket{e}\Big) \otimes \ket{0_M}. \label{eq:resonant_final}
\end{align}
\end{widetext}

\begin{figure}[h!]
    \centering
    \includegraphics[width=\linewidth]{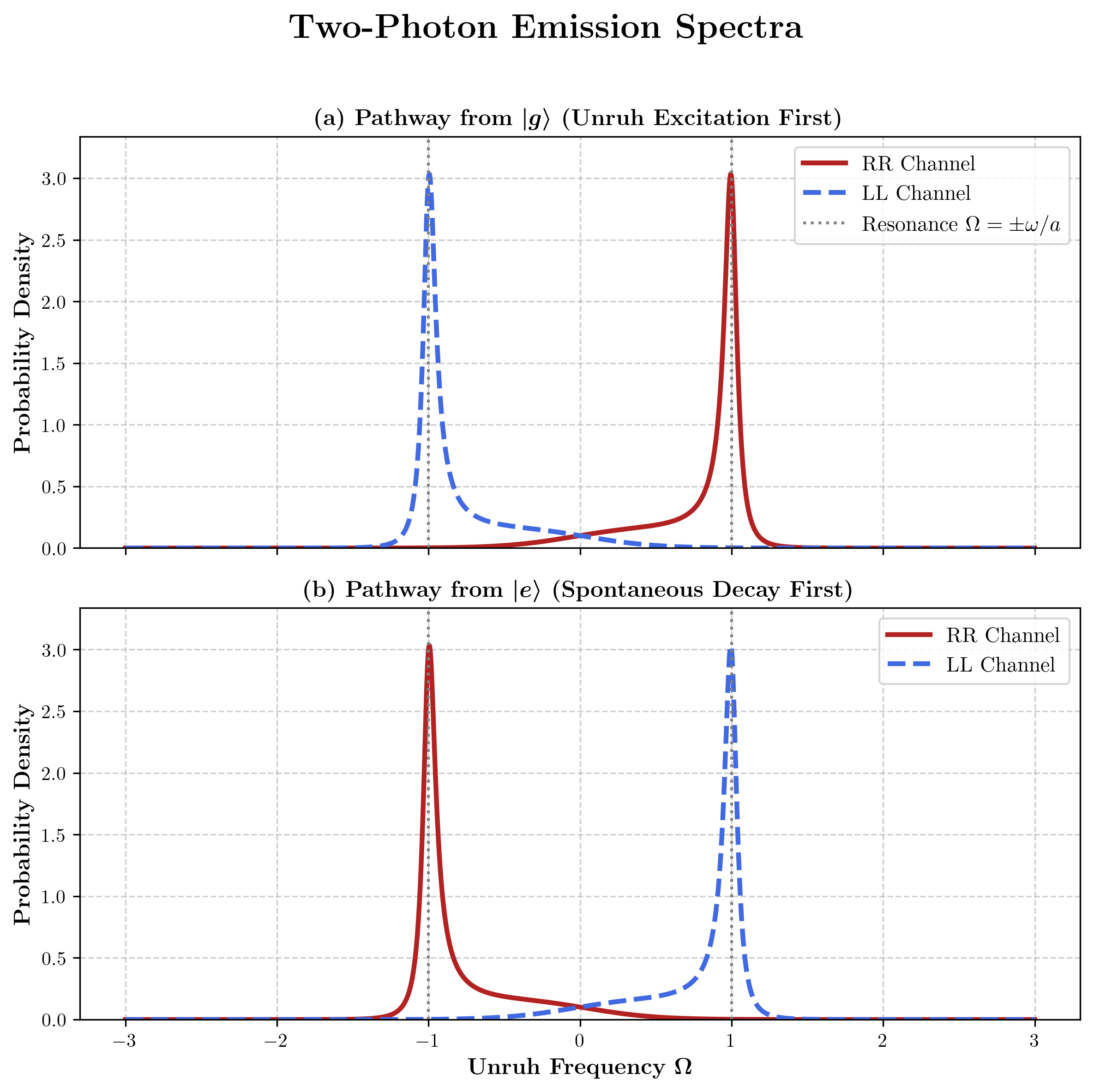}
   \caption{Two-photon emission spectra for the two pathways. (a) For the $g \to e \to g$ process, the RR channel is resonant at positive $\Omega$ and the LL channel at negative $\Omega$. (b) For the $e \to g \to e$ process, the roles are inverted, with the LL channel resonant at positive $\Omega$ and the RR at negative $\Omega$.}
    \label{fig:spectra}
\end{figure}

\textit{Emission Spectra and Directional Control.}---The two-photon emission spectra, shown in Fig.~\ref{fig:spectra}, provide a clear visualization of the directional control exerted by the detector's initial state. The plot reveals that for each pathway, the RR and LL channels are resonantly enhanced at opposite frequency signs, as summarized in Table~\ref{tab:spectra_summary_revised}.

In the $g \to e \to g$ process, the RR channel emission is overwhelmingly dominant at positive frequencies, with its sharp resonant peak located at $\Omega = +\omega/a$, while the LL channel is dominant at negative frequencies, with its peak at $\Omega = -\omega/a$. Conversely, in the $e \to g \to e$ process, the RR channel is dominant at negative frequencies ($\Omega = -\omega/a$), and the LL channel is dominant at positive frequencies ($\Omega = +\omega/a$). As a direct consequence of the underlying $\omega \to -\omega$ symmetry, the peak height of the RR emission at positive frequencies in the first pathway is identical to that of the LL emission at positive frequencies in the second pathway. This demonstrates that the initial state acts as a switch, coherently controlling which directional channel is dominantly populated at a given frequency sign.

\begin{table}[h!]
    \centering
    \caption{Summary of emission channels. For each pathway, channels are dominantly enhanced at specific frequency signs.}
    \vspace{0.2cm}
    \renewcommand{\arraystretch}{1.5}
    \begin{tabular}{lcc}
        \toprule
        \textbf{Initial Pathway} & \textbf{RR Dominant At} & \textbf{LL Dominant At} \\
        \midrule
        $|g\rangle \to |e\rangle \to |g\rangle$ & Positive $\Omega$ ($+\omega/a$) & Negative $\Omega$ ($-\omega/a$) \\
        $|e\rangle \to |g\rangle \to |e\rangle$ & Negative $\Omega$ ($-\omega/a$) & Positive $\Omega$ ($+\omega/a$) \\
        \bottomrule
    \end{tabular}
    \label{tab:spectra_summary_revised}
\end{table}


\begin{figure}[h!]
    \centering
    \includegraphics[width=\linewidth]{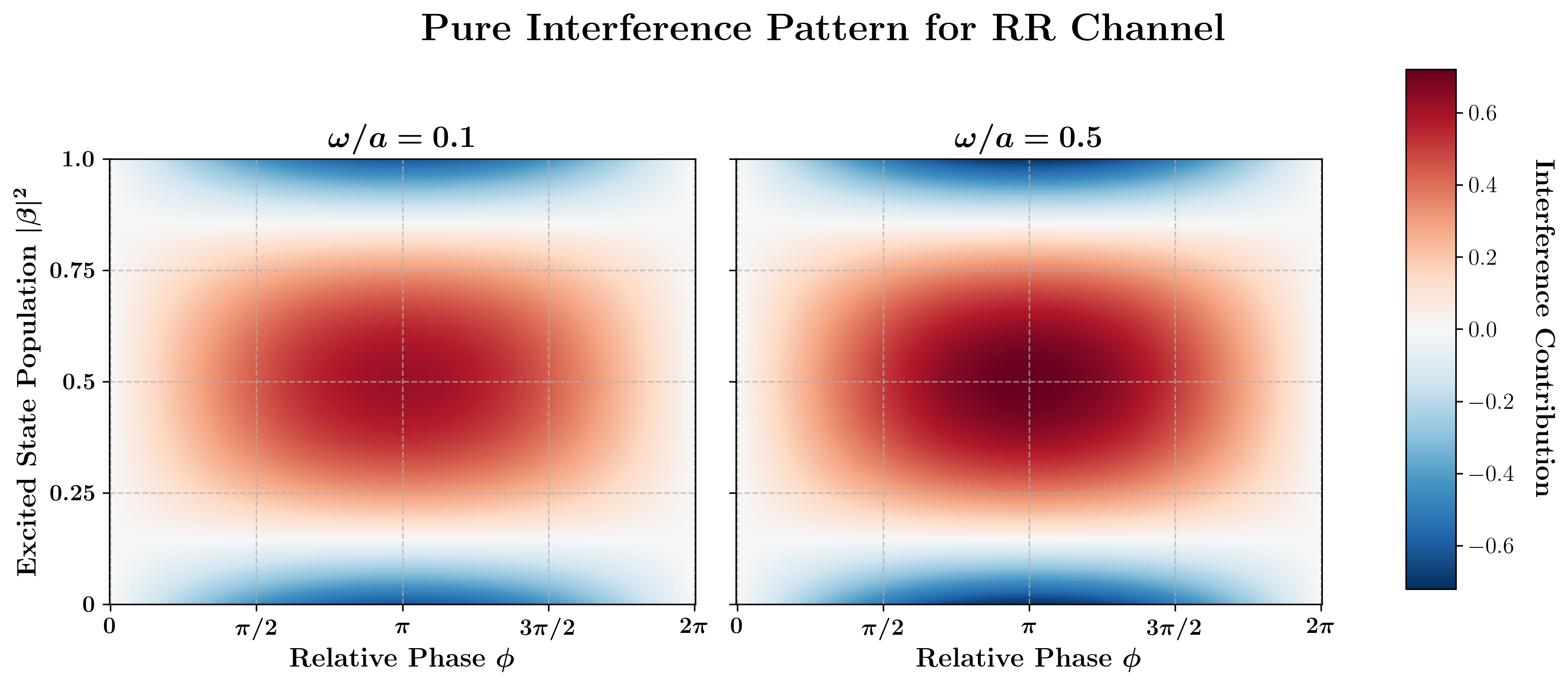}
    \caption{The pure interference contribution to the RR channel emission probability. The color indicates constructive (red) or destructive (blue) interference.}
    \label{fig:rr_interference}
\end{figure}

\begin{figure}[h!]
    \centering
    \includegraphics[width=\linewidth]{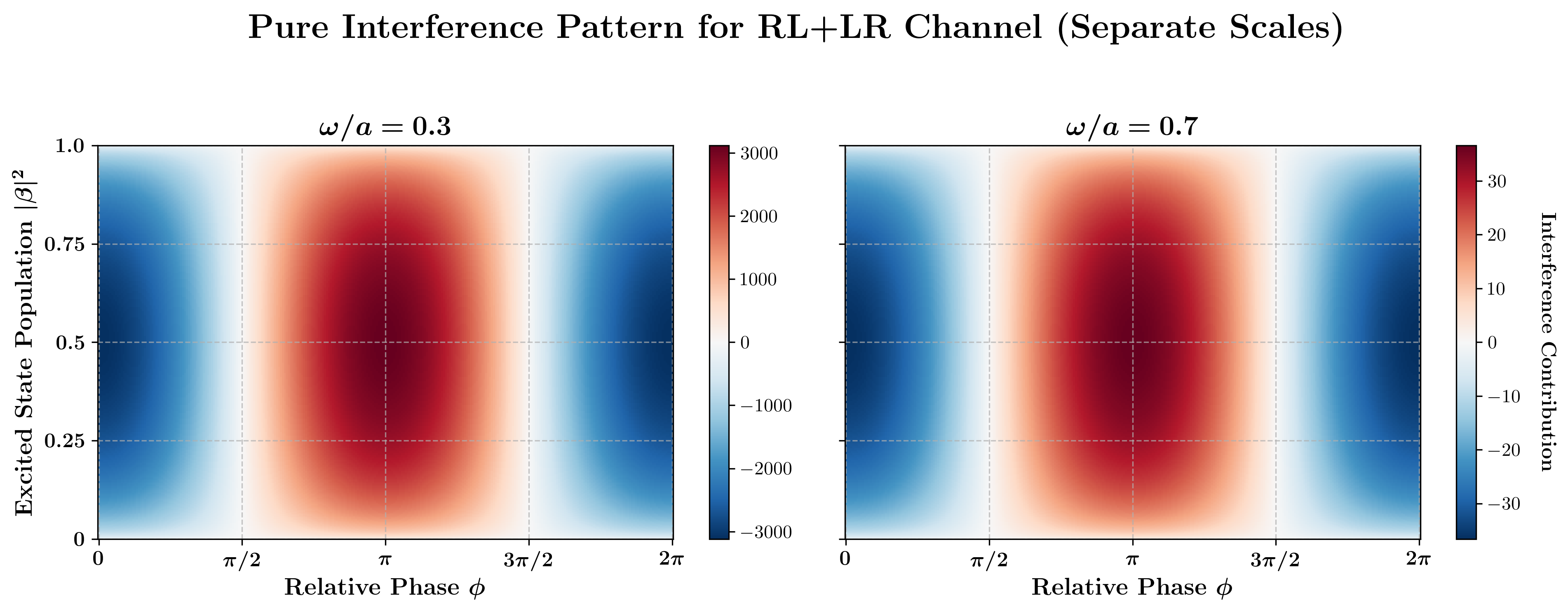}
    \caption{The pure interference contribution to the RL+LR channel, plotted with separate color scales. The pattern is inverted compared to the RR channel and its magnitude drops dramatically with increasing $\omega/a$.}
    \label{fig:rl_interference}
\end{figure}

\textit{Numerical Results and Visualization.}---To explore the coherent control enabled by preparing the detector in a superposition, we analyze the interference between the two pathways. The total emission probability can be decomposed as $P_{\text{total}} = P_{\text{background}} + P_{\text{int}}$, where $P_{\text{background}}$ is the incoherent sum and $P_{\text{int}}$ is the pure interference term. The color in the heatmaps of Figs.~\ref{fig:rr_interference} and \ref{fig:rl_interference} represents this interference contribution, $P_{\text{int}}$. A positive (red) value signifies constructive interference, while a negative (blue) value signifies destructive interference where the emission is suppressed.

Figure~\ref{fig:rr_interference} shows the interference pattern for the RR channel. Constructive interference is maximal for an equal superposition ($|\beta|^2=0.5$) with a relative phase of $\phi=\pi$. The interference strength is also non-monotonic with the ratio $\omega/a$, revealing a ``sweet spot'' that maximizes visibility. We note that the total integrated probability for the LL channel is identical to that of the RR channel due to an underlying integration symmetry ($\Omega \to -\Omega$), meaning their interference heatmaps are the same. In contrast, Fig.~\ref{fig:rl_interference} shows the pattern for the mixed RL+LR channel is inverted, with constructive interference now at $\phi=0$. Most strikingly, the magnitude of this mixed-channel interference plummets as $\omega/a$ increases, demonstrating that it is a distinct feature of the low-energy-gap, high-acceleration regime.

Finally, to provide an unambiguous signature of the quantum nature of the emitted radiation, we analyze its phase-space properties. The final state is a two-photon state, entangled across different modes (RR, LL, etc.). To characterize a single photon, we compute its reduced density matrix by tracing over its entangled partner. We then calculate the Wigner function for this single-photon reduced state.

\begin{figure}[h!]
    \centering
    \includegraphics[width=\linewidth]{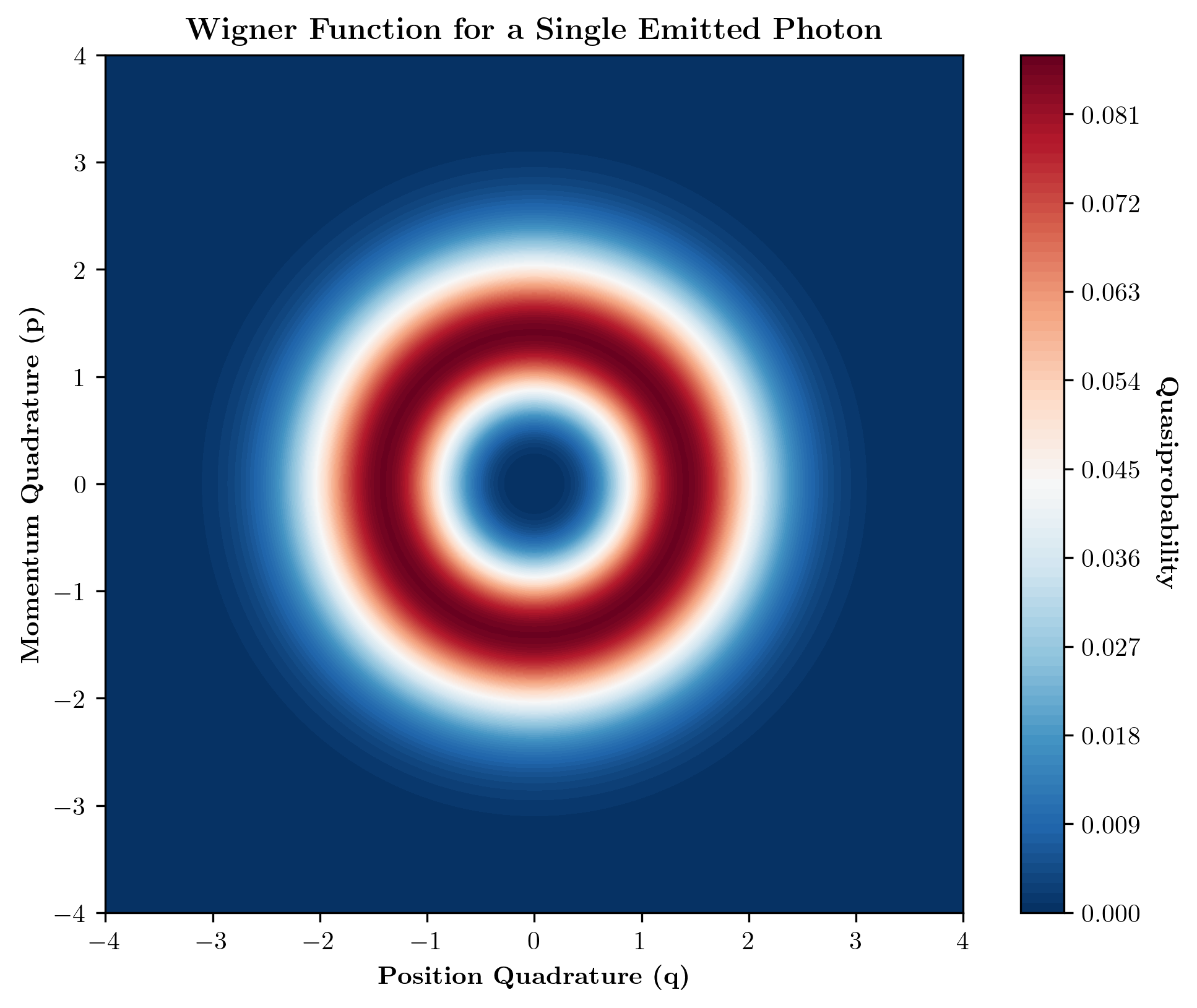}
    \caption{A 2D plot of the Wigner function for a single emitted photon. The central negative (blue) region is a definitive signature of the state's non-classicality.}
    \label{fig:wigner2d}
\end{figure}

\begin{figure}[h!]
    \centering
    \includegraphics[width=\linewidth]{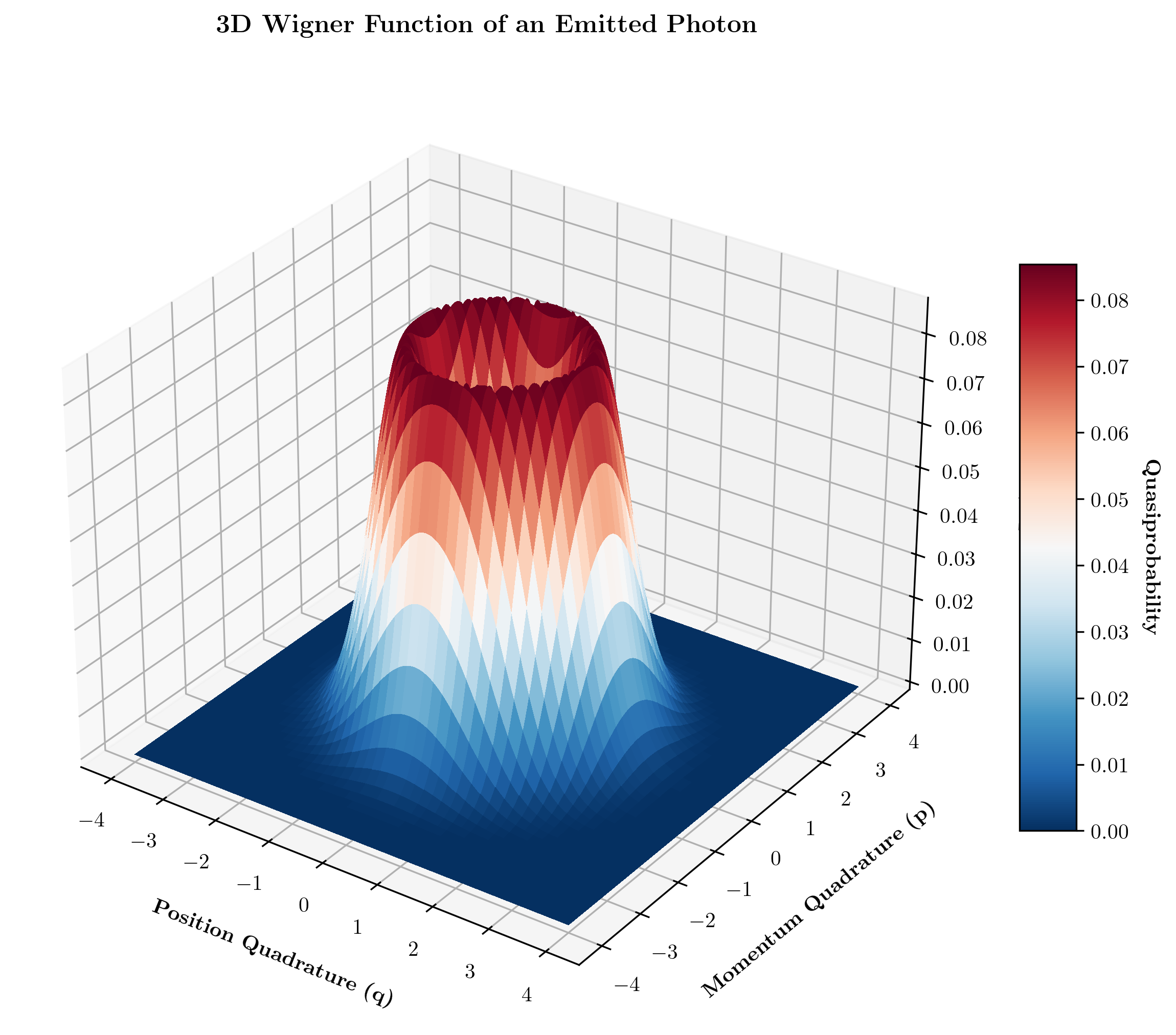}
    \caption{A 3D surface plot of the Wigner function shown in Fig.~\ref{fig:wigner2d}, providing an intuitive view of the negative quasiprobability at the origin of phase space.}
    \label{fig:wigner3d}
\end{figure}

Figures~\ref{fig:wigner2d} and \ref{fig:wigner3d} show the resulting Wigner function. Its most crucial feature is the prominent negative region at the origin of phase space. A classical state of light, including a thermal state like that typically associated with the Unruh effect, must have a non-negative Wigner function. This negativity is therefore a definitive, ``smoking gun'' proof of the non-classical character of the emitted radiation. It demonstrates that the vacuum-induced process does not merely produce thermal noise, but generates a deeply quantum state with no classical analogue.

\textit{Experimental Feasibility.}---While direct observation of the two-photon Unruh effect is beyond current technology due to the immense accelerations required, the quantum gate we describe is well-suited for verification in analog quantum systems~\cite{Barcelo2011LRR}. Platforms such as circuit QED~\cite{Nation2012Nori, Wilson2011Nature}, trapped ions~\cite{Alsing2005IonTrap}, or Bose-Einstein condensates~\cite{Steinhauer2016} have been proposed for simulating relativistic quantum phenomena, and the core components of our proposal are within reach of current experimental capabilities.

A promising platform is circuit QED, where a superconducting transmon qubit acts as the UDW detector and the electromagnetic field in a transmission line resonator plays the role of the quantum field. The effect of uniform acceleration can be simulated by rapidly modulating system parameters, such as the qubit's frequency or the boundary conditions of the resonator, to mimic the time-dependent coupling experienced by an accelerating observer~\cite{Nation2012Nori}. This technique is closely related to the experimental observation of the dynamical Casimir effect, which generates real photons from the vacuum via moving mirrors~\cite{Wilson2011Nature}.

The primary experimental signature would be the verification of the conditional phase flip on the detector's state. This can be measured directly using Ramsey interferometry, a technique already proposed for measuring the Unruh effect~\cite{Polo-Gomez2022measurement}:
\begin{enumerate}
    \item \textbf{Preparation:} A Hadamard-like gate prepares the qubit in an equal superposition state $\frac{1}{\sqrt{2}}(\ket{g} + \ket{e})$.
    \item \textbf{Interaction:} The time-dependent modulation simulating the acceleration is applied, allowing the second-order interaction with the vacuum field to occur.
    \item \textbf{Measurement:} A second Ramsey pulse is applied to the qubit, followed by a projective measurement of its state.
\end{enumerate}
The outcome of the measurement will depend on the relative phase acquired by the qubit during the interaction. The vacuum-induced gate mechanism predicts a $\pi$ phase shift, which would manifest as an inversion of the Ramsey interference fringes. Observing this fringe inversion would provide conclusive evidence of the coherent, computational nature of the interaction. Furthermore, advanced measurements could probe the correlations of the emitted microwave photons in the resonator to verify the two-photon nature of the process, providing a complete experimental confirmation of the theory.

\textit{Conclusion.}---The resonant final state in Eq.~\eqref{eq:resonant_final} reveals the core of our findings. The interaction transforms the detector's initial state $\alpha\ket{g} + \beta\ket{e}$ into $\alpha\ket{g} - \beta\ket{e}$, which is precisely the operation of a quantum Z-gate. The name ``vacuum-induced quantum gate'' is justified because the entire process is driven by the structure of the quantum vacuum as perceived by the accelerating detector. The two-photon emission, which heralds the successful operation of the gate, is a direct result of harvesting entanglement from the Minkowski vacuum across causally disconnected Rindler wedges.

Most significantly, we have shown that the relativistic vacuum, when probed correctly, functions as a coherent computational resource. This work reframes the Unruh effect not as a source of thermal noise to be mitigated, but as a fundamental mechanism that can be harnessed for quantum information processing, opening new pathways at the intersection of quantum gravity and quantum computing.

\textit{Acknowledgments.}---I am grateful to Girish Agarwal, Marlan Scully, Bill Unruh, and Suhail Zubairy for discussions. This work was supported by the
Robert A. Welch Foundation (Grant No. A-1261) and the National Science Foundation (Grant No. PHY-2013771).

\bibliographystyle{apsrev4-2}
\bibliography{UnruhRef}


\newpage
\clearpage
\onecolumngrid
\appendix
\section*{Supplemental Material for ``Vacuum-Induced Quantum Gate''}
\author{Arash Azizi}
\affiliation{The Institute for Quantum Science and Engineering, Texas A\&M University, College Station, TX 77843, U.S.A.}
\affiliation{Department of Physics and Astronomy, Texas A\&M University, College Station, TX 77843, U.S.A.}
\date{\today}
\maketitle

In this Supplemental Material, we provide the detailed derivations supporting the results presented in the main Letter.
\setcounter{equation}{0}
\renewcommand{\theequation}{S\arabic{equation}}

\section{Field Quantization and Unruh Modes} \label{sec:field_quantization}

In this section, we report the essential formulas for the Unruh-DeWitt (UDW) detector model and the Unruh mode expansion of the scalar field. This material is included to make the Supplemental Material self-contained and to establish our notation.

We consider a two-level UDW detector with energy gap $\omega$ following a path of uniform proper acceleration $a$. The interaction with a massless scalar field $\Phi$ in (1+1) dimensions is described by the Hamiltonian:
\begin{equation}
H_{\text{int}} = g \frac{\partial }{\partial \tau} \Phi(x(\tau))
\left( \sigma^{\dagger} e^{i \omega \tau} + \sigma e^{-i \omega \tau} \right). \label{H_int}
\end{equation}
The detector's trajectory is given by the light-cone coordinates $u = -\frac{1}{a} e^{-a\tau}$ and $v = \frac{1}{a} e^{a\tau}$. The scalar field is decomposed into right-moving ($\Phi_{\text{RTW}}$) and left-moving ($\Phi_{\text{LTW}}$) components, which are expanded in the Unruh mode basis as:
\begin{align}
\Phi_{\text{RTW}}(u)= \int_{-\infty}^{+\infty} d\Omega
\Bigg\{& \Big(
\theta(u) f(\Omega) u^{i\Omega}
+ \theta(-u)f(-\Omega) (-u)^{i\Omega} \Big) A_{\Omega} \nn\\
& + \Big(
\theta(u) f(\Omega) u^{-i\Omega}
+ \theta(-u) f(-\Omega) (-u)^{-i\Omega} \Big) A_{\Omega}^{\dagger}
\Bigg\}, \label{Unruh.mode.RTW}
\end{align}
and
\begin{align}
\Phi_{\text{LTW}}(v)= \int_{-\infty}^{+\infty} d\Omega
\Bigg\{& \Big(
\theta(v) f(\Omega) v^{i\Omega}
+ \theta(-v)f(-\Omega) (-v)^{i\Omega} \Big) B_{\Omega} \nn\\
& + \Big(
\theta(v) f(\Omega) v^{-i\Omega}
+ \theta(-v) f(-\Omega) (-v)^{-i\Omega} \Big) B_{\Omega}^{\dagger}
\Bigg\}, \label{Unruh.mode.LTW}
\end{align}
where $A_{\Omega}$ and $B_{\Omega}$ are the annihilation operators for right- and left-moving Unruh modes. The normalization coefficient is given by
\begin{equation}
f(\Omega) = \frac{e^{- \pi \Omega / 2}}{\sqrt{8\pi \Omega \sinh(\pi \Omega)}}.\label{f}
\end{equation}

The detector's path is parameterized as $u = -\frac{1}{a} e^{-a\tau}$ and $v = \frac{1}{a} e^{a\tau}$. In the detector's proper time $\tau$ within the right Rindler wedge ($u < 0, v > 0$), the field operators become:
\begin{align}
\Phi_{\text{RTW}}(\tau)
&= \int_{-\infty}^{+\infty} d\Omega f(-\Omega)\Big(
a^{-i\Omega} e^{-i a \Omega \tau} A_\Omega
+ a^{i\Omega} e^{i a \Omega \tau} A_\Omega^\dagger
\Big), \nn \\
\Phi_{\text{LTW}}(\tau)
&= \int_{-\infty}^{+\infty} d\Omega
f(\Omega)\Big(
a^{-i\Omega} e^{i a \Omega \tau} B_\Omega
+ a^{i\Omega} e^{-i a \Omega \tau} B_\Omega^\dagger \Big). \label{Unruh.mode.tau}
\end{align}
The Jacobian transformations
\begin{equation}
du \frac{\partial}{\partial u} = d\tau \frac{\partial}{\partial \tau}, \quad dv \frac{\partial}{\partial v} = d\tau \frac{\partial}{\partial \tau},
\end{equation}
facilitate the conversion of the Dyson series to integrals over proper time.
\vspace{.3cm}

These definitions are sufficient to proceed with the second-order calculations that form the basis of our work.

\section{Two-Photon Emission from an Excited Detector} \label{sec. time evol}

We first calculate the final two-photon state for a detector that begins in its excited state $\ket{e}$ with the field in the Minkowski vacuum $\ket{0_M}$. The transition pathway is $e \to g \to e$, corresponding to a radiative decay followed by an Unruh re-excitation. The state evolution is governed by the second-order Dyson series:
\begin{align}
\ket{\Psi_{e \to g \to e}} = \left( \frac{-i}{\hbar} \right)^2
\int_{-\infty}^{+\infty} d\tau H_{\text{int}}(\tau)
\int_{-\infty}^{\tau} d\tau' H_{\text{int}}(\tau')
\ket{0_M} \ket{e}.
\label{psi_f-tau_setup}
\end{align}
To describe the $e \to g \to e$ process, the operator at the earlier time $\tau'$ must cause a decay ($\sigma e^{-i\omega\tau'}$), and the operator at the later time $\tau$ must cause an excitation ($\sigma^{\dagger} e^{i\omega\tau'}$). The correctly time-ordered state is therefore:
\begin{align}
\ket{\Psi_{e \to g \to e}} = \left( \frac{-i}{\hbar} \right)^2
\int_{-\infty}^{+\infty} d\tau g \frac{\partial \Phi}{\partial \tau} \sigma^{\dagger} e^{i \omega \tau}
\int_{-\infty}^{\tau} d\tau' g \frac{\partial \Phi}{\partial \tau'} \sigma e^{-i \omega \tau'}
\ket{0_M} \ket{e}.
\label{psi_f-tau}
\end{align}
Since the field begins in the vacuum, only the creation parts of the field operators from Eq. \eqref{Unruh.mode.tau} contribute. The calculation involves performing the nested time integrals. The inner integral over $\tau'$ produces the energy denominator corresponding to the intermediate state (detector in state $\ket{g}$), while the outer integral over $\tau$ enforces overall energy conservation between the two Unruh photons and the two detector transitions, yielding a Dirac delta function.

After evaluating these integrals for all possible combinations of right-moving ($A^\dagger$) and left-moving ($B^\dagger$) photons, we arrive at the final analytical form of the two-photon state:
\begin{align}
 \ket{\Psi_{e \to g \to e}} =
\frac{ -i g^2}{4\hbar^2}
\int_{-\infty}^{+\infty} d\Omega
\frac{1}{\sinh(\pi \Omega)} 
\Bigg\{ \frac{\Omega}{\frac{\omega}{a} + \Omega}
 A^\dagger_{\Omega}  A^\dagger_{-\Omega}
 + \frac{\Omega}{\frac{\omega}{a} - \Omega}
 B^\dagger_{\Omega}  B^\dagger_{-\Omega} 
 + \frac{2\frac{\omega}{a}\,\Omega\,  a^{2i\Omega} }{(\frac{\omega}{a})^2 - \Omega^2}
 A^\dagger_{\Omega}  B^\dagger_{\Omega} \Bigg\}
\ket{0_M} \ket{e}. \label{ege}
\end{align}
This state describes the correlations between the emitted photons. The denominators reveal the resonant structure of the emission. For instance, the Left-Left (LL) channel, with its denominator $(\omega/a - \Omega)$, is resonant when the Unruh frequency of one emitted photon matches the detector's transition frequency, $\Omega = \omega/a$. This corresponds physically to the detector decaying ($e \to g$) by emitting an L-moving Unruh photon of frequency $\omega/a$, followed by an Unruh excitation ($g \to e$) that creates a correlated L-moving photon of frequency $-\omega/a$.

\section{Symmetry Analysis and Interference Effects}
\subsection{The Counterpart Process: \texorpdfstring{$g \to e \to g$}{g -> e -> g}}
To fully understand the physics of the $e \to g \to e$ transition, it is illuminating to compare it with its counterpart process: two-photon emission from a detector initially in the ground state, $\ket{g}$. This process follows the pathway $g \to e \to g$, where the detector is first excited by the Unruh effect and subsequently decays back to the ground state. We report the complete answer here:
\begin{align}
\ket{\Psi_{g \to e \to g}} =
\frac{ i g^2}{4\hbar^2}
\int_{-\infty}^{+\infty} d\Omega
\frac{1}{\sinh(\pi \Omega)} 
\Bigg\{ \frac{\Omega}{\frac{\omega}{a} - \Omega}
 A^\dagger_{\Omega}  A^\dagger_{-\Omega}
 + \frac{\Omega}{\frac{\omega}{a} + \Omega}
 B^\dagger_{\Omega}  B^\dagger_{-\Omega} + \frac{2\frac{\omega}{a}\,\Omega\,  a^{2i\Omega} }{(\frac{\omega}{a})^2 - \Omega^2}
 A^\dagger_{\Omega}  B^\dagger_{\Omega} \Bigg\}
\ket{0_M} \ket{g}. \label{geg}
\end{align}

\subsection{The \texorpdfstring{$\omega \to -\omega$}{w -> -w} Symmetry}
By comparing the final states in Eq. \eqref{ege} and Eq. \eqref{geg}, a profound symmetry becomes apparent. The internal evolution of the detector between the two interactions (from $\tau'$ to $\tau$) involves a phase factor $e^{+i\omega(\tau-\tau')}$ for the $e \to g \to e$ process and $e^{-i\omega(\tau-\tau')}$ for the $g \to e \to g$ process. This is the only difference in the underlying calculation, and it is equivalent to transforming the detector's energy gap $\omega \to -\omega$. This leads to a direct relationship between the two transition amplitudes ($\mathcal{A}$):
\begin{equation}
    \mathcal{A}_{e \to g \to e}(\omega) = \mathcal{A}_{g \to e \to g}(-\omega)
\end{equation}
This symmetry provides a powerful conceptual check on our results and reveals a fundamental connection between particle creation processes that begin with either spontaneous decay or Unruh excitation.

\subsection{Superposition}

The evolution of the quantum system is linear. Therefore, given an initial superposition state
\begin{align}
    \ket{\Psi_i} = \left( \alpha \ket{g} + \beta \ket{e} \right) \otimes \ket{0_M},
\end{align}
the final state $\ket{\Psi_f}$ is the superposition of the final states evolving from each component independently:
\begin{align}
    \ket{\Psi_f} = \alpha \ket{\Psi_{g \to e \to g}}
    + \beta \ket{\Psi_{e \to g \to e}} .
\end{align}
Substituting the explicit expressions for each pathway, we obtain the final entangled state of the detector-field system. We can combine the terms under a single integral by factoring out the common field creation operators:

\begin{empheq}[box=\widefbox]{align}
\ket{\Psi_f} = \frac{i g^2}{4\hbar^2} \int_{-\infty}^{+\infty} &d\Omega \frac{1}{\sinh(\pi \Omega)} \Bigg\{ \nn \\
&\alpha \Bigg( \frac{\Omega}{\frac{\omega}{a} - \Omega} A^\dagger_{\Omega} A^\dagger_{-\Omega} + \frac{\Omega}{\frac{\omega}{a} + \Omega} B^\dagger_{\Omega} B^\dagger_{-\Omega} + \frac{2\frac{\omega}{a}\,\Omega\, a^{2i\Omega} }{(\frac{\omega}{a})^2 - \Omega^2} A^\dagger_{\Omega} B^\dagger_{\Omega} \Bigg) \ket{0_M} \otimes \ket{g} \nn \\
- &\beta \Bigg( \frac{\Omega}{\frac{\omega}{a} + \Omega} A^\dagger_{\Omega} A^\dagger_{-\Omega} + \frac{\Omega}{\frac{\omega}{a} - \Omega} B^\dagger_{\Omega} B^\dagger_{-\Omega} + \frac{2\frac{\omega}{a}\,\Omega\, a^{2i\Omega} }{(\frac{\omega}{a})^2 - \Omega^2} A^\dagger_{\Omega} B^\dagger_{\Omega} \Bigg) \ket{0_M} \otimes \ket{e} \Bigg\}. \label{eq:fully_correct_entangled_state}
\end{empheq}

\section{Evaluating the Final State Using the Sokhotski--Plemelj Theorem}

The final entangled state $\ket{\Psi_f}$ is given by the integral expression in Eq.~\eqref{eq:fully_correct_entangled_state}. To evaluate this integral, which contains singularities in the denominators at $\Omega = \pm \frac{\omega}{a}$, we employ the Sokhotski--Plemelj theorem. This theorem allows us to handle the principal value integrals and extract the delta-function contributions arising from the poles on the real axis.

Let us denote $\Omega_0 = \frac{\omega}{a}$ for brevity. The integral is over $\Omega \in (-\infty, +\infty)$, and we regularize the denominators with a small positive imaginary part $\epsilon > 0$, taking the limit $\epsilon \to 0^+$ at the end. The Sokhotski--Plemelj theorem states that
\[
\lim_{\epsilon \to 0^+} \frac{1}{x + i \epsilon} = \mathcal{P} \frac{1}{x} - i \pi \delta(x),
\]
where $\mathcal{P}$ denotes the Cauchy principal value.

We apply this to each term in the integrand. Consider a general term of the form
\[
\int_{-\infty}^{+\infty} d\Omega \, \frac{h(\Omega)}{\Omega_0 - \Omega},
\]
where $h(\Omega) = \frac{\Omega}{\sinh(\pi \Omega)} \, \hat{O}(\Omega)$ and $\hat{O}(\Omega)$ represents the corresponding operator acting on the state (e.g., $A^\dagger_{\Omega} A^\dagger_{-\Omega} \ket{0_M} \otimes \ket{g}$). Regularizing as
\[
\int_{-\infty}^{+\infty} d\Omega \, \frac{h(\Omega)}{\Omega_0 - \Omega + i \epsilon} = \mathcal{P} \int_{-\infty}^{+\infty} d\Omega \, \frac{h(\Omega)}{\Omega_0 - \Omega} - i \pi h(\Omega_0).
\]

Similarly, for terms with denominator $\Omega_0 + \Omega$,
\[
\int_{-\infty}^{+\infty} d\Omega \, \frac{h(\Omega)}{\Omega_0 + \Omega + i \epsilon} = \mathcal{P} \int_{-\infty}^{+\infty} d\Omega \, \frac{h(\Omega)}{\Omega_0 + \Omega} - i \pi h(-\Omega_0).
\]

For the third term, $\frac{2 \Omega_0\, \Omega\, a^{2 i \Omega}}{\Omega_0^2 - \Omega^2} = \Omega_0 \left[ \frac{1}{\Omega_0 - \Omega} - \frac{1}{\Omega_0 + \Omega} \right] a^{2 i \Omega}$, we decompose it using partial fractions and apply the regularization to each part separately:
\[
\frac{1}{\Omega_0 - \Omega + i \epsilon} - \frac{1}{\Omega_0 + \Omega + i \epsilon} = \mathcal{P} \left( \frac{1}{\Omega_0 - \Omega} - \frac{1}{\Omega_0 + \Omega} \right) - i \pi \left[ \delta(\Omega_0 - \Omega) - \delta(\Omega_0 + \Omega) \right].
\]
Since,
\[
\frac{\Omega}{\Omega_0 - \Omega} = \Omega_0 (\Omega_0 - \Omega)^{-1} - 1, \quad \frac{\Omega}{\Omega_0 + \Omega} = - \Omega_0 (\Omega_0 + \Omega)^{-1} + 1,
\]
then the non-singular $-1$ or $+1$ terms lead to regular integrals, while the singular parts are handled via the theorem.

We apply this to each singular term in the integrand. The state splits into
\[
\ket{\Psi_f} = \ket{\Psi_f}^{\rm PV} + \ket{\Psi_f}^{\rm res},
\]
where $\ket{\Psi_f}^{\rm PV}$ is the principal value contribution,
\begin{align}
    \ket{\Psi_f}^{\rm PV} =& \frac{i g^2}{4\hbar^2} \mathcal{P} \int_{-\infty}^{+\infty} d\Omega \frac{1}{\sinh(\pi \Omega)} \Bigg\{ \alpha \Bigg( \frac{\Omega}{\Omega_0 - \Omega} A^\dagger_{\Omega} A^\dagger_{-\Omega} + \frac{\Omega}{\Omega_0 + \Omega} B^\dagger_{\Omega} B^\dagger_{-\Omega} + \frac{2\Omega_0\,\Omega\, a^{2i\Omega} }{\Omega_0^2 - \Omega^2} A^\dagger_{\Omega} B^\dagger_{\Omega} \Bigg) \ket{g} \nn\\
&- \beta \Bigg( \frac{\Omega}{\Omega_0 + \Omega} A^\dagger_{\Omega} A^\dagger_{-\Omega} + \frac{\Omega}{\Omega_0 - \Omega} B^\dagger_{\Omega} B^\dagger_{-\Omega} + \frac{2\Omega_0\,\Omega\, a^{2i\Omega} }{\Omega_0^2 - \Omega^2} A^\dagger_{\Omega} B^\dagger_{\Omega} \Bigg)   \ket{e} \Bigg\}\ket{0_M},
\end{align}
and $\ket{\Psi_f}^{\rm res}$ is the resonant (delta-function) contribution.

To compute $\ket{\Psi_f}^{\rm res}$, we collect the $-i\pi \delta$ terms from each pole. After decomposing the terms and applying the theorem, the resonant contribution simplifies to
\[
\ket{\Psi_f}^{\rm res} = \frac{i g^2}{4 \hbar^2} \, i \gamma \left( \beta \, O \ket{0_M} \otimes \ket{e} - \alpha \, O \ket{0_M} \otimes \ket{g} \right) = -\frac{g^2 \gamma}{4 \hbar^2} \left( \beta \, \ket{e} - \alpha \, \ket{g} \right) \otimes O \ket{0_M},
\]
where
\[
\gamma = \frac{\pi \Omega_0}{\sinh(\pi \Omega_0)},
\]
and
\[
O = A^\dagger_{\Omega_0} A^\dagger_{-\Omega_0} + B^\dagger_{\Omega_0} B^\dagger_{-\Omega_0} + a^{2i \Omega_0} A^\dagger_{\Omega_0} B^\dagger_{\Omega_0} + a^{-2i \Omega_0} A^\dagger_{-\Omega_0} B^\dagger_{-\Omega_0}.
\]
This resonant part captures the on-shell contributions, physically corresponding to real processes such as particle pair creation or emission at the resonance frequency $\Omega_0$. The principal value part $\ket{\Psi_f}^{\rm PV}$ represents off-resonant virtual processes and may require numerical evaluation or further analytical techniques (e.g., contour integration using the partial fraction expansion of $\csch(\pi \Omega)$) for explicit computation, depending on the specific model details. The final answer reads:

\begin{empheq}[box=\widefbox]{align}
\ket{\Psi_f}_{\text{res}} \approx \frac{\pi g^2}{4\hbar^2} \frac{\omega/a}{\sinh(\pi \omega/a)} \Bigg\{& A^\dagger_{\omega/a} A^\dagger_{-\omega/a} 
+ B^\dagger_{\omega/a} B^\dagger_{-\omega/a}  \nn\\
& + a^{2i\omega/a} A^\dagger_{\omega/a} B^\dagger_{\omega/a} 
+ a^{-2i\omega/a} A^\dagger_{-\omega/a} B^\dagger_{-\omega/a}    \Bigg\} \Big(\alpha\ket{g} - \beta\ket{e}\Big)  \otimes \ket{0_M}.
\end{empheq}

\end{document}